\documentclass{llncs}
\usepackage{graphicx}
\usepackage{subfig}
\usepackage{amsmath}
\usepackage{amssymb}
\usepackage[T1]{fontenc}
\usepackage[utf8]{inputenc}

\usepackage{todonotes}

\begin{document}

\title{From close the door to do not click and back. Security by design for older adults.}
%
%
\author{Bartlomiej Balcerzak, Wieslaw Kopec, Radoslaw Nielek, Kamil Warpechowski, Agnieszka Czajka}
\institute{Polish-Japanese Institute of Information Technology,\\ 
              ul. Koszykowa 86, 02-008 Warsaw, Poland\\}


\maketitle              

\begin{abstract}
With the growing number of older adults who adopt mobile technology in their life, a new form of challenge faces both them, as well as the software engineering communities. This challenge is the issue of safety, not only in the context of risk older adults already face on-line, but also, due to the mobile nature of the used applications, real life safety issues raising from the use of on-line solutions. In this paper, we wish to use a case study they conducted in order to address this issue of interrelating on-line and real life threats. We describe how the observation from the case study relate to the collected body off knowledge in the relevant topic, as well as propose a set of suggestion for improving the design of applications in regards to addressing the issue of older adults safety.
\end{abstract}

\section{Introduction}

The trend of population aging is currently an undisputed fact, observed throughout the globe, in developed and developing countries alike \cite{nations2015world}. By 2050 over 20\% of the population of the USA is projected to be 65+ \cite{ortman2014aging}, other developed countries, such as Japan are also expected to become aging populations by that time. Although this trend has been deemed a negative trajectory for population development \cite{harper2014economic}, and a giant challenge for social security frameworks, and in turn a major burden for the state, research shows that this trend may bear out other results.  For instance, Spijker et al. argument that although demographic data are true in reality number of older adults requiring assistance in UK and other countries actually been falling in recent years \cite{spijker2013population}. This point of view is supported by Sanderson et al. \cite{sanderson2015faster} who point out that metrical age qualifying to being older adult will increase in coming years.

In parallel to the mentioned trend, an increase in smartphone and on-line tool usage among older adults can also be observed \cite{zickuhr2012older}. This trend of increasing usage of mobile ICT can be observed worldwide, and already has inspired many researchers and designers to introduce new and creative solutions tailored towards the needs of the older adults \cite{massimi2007using}. When considered in relation to the findings suggesting that in future aging societies, older adults may be more affluent, healthy and tech savvy than their current counterparts. This increase in use of mobile technology, opens up new possibilities for addressing the issue of population aging, and creating solutions tailored to a new growing customer base. With these new possibilities however, also come new challenges. The demographics increase in the share of mobile solutions users leads to new, and previously not addressed issues such as those related to the older adults safety.

Research shows that older adult are less likely to be tech savvy\cite{grimes2010older}, and often as a result, are more prone to suffer from threats related to their safety. Moreover, older adults tend to be more aware of the risks they face on-line \cite{adams2005psychological}. Due to the fact that on-line solutions such as social media, communication and financial services or other mobile based apps increasingly blur the line between the on-line and real life situation, it is understandable that, when addressing the safety issues of older adults, one must consider both aspects of safety. Discussing this topic is our main aim in this paper.

For the sake of this paper, we made an explicit distinction between two aspects of older adults safety; their on-line security, and real life safety. The former, as the name suggest, refers to risk the older adult may face when interacting with ICT (ie. identity theft, scamming, phishing, spread of malware and spam), the latter on the other hand deals with risks connected with interactions made in the real world, such as theft, assault, burglary, traffic accidents on so on. Although these two aspects were deeply reviewed by researchers (with work by \cite{grimes2010older}, and \cite{forjuoh2016172} serving as good examples), there has been no attempt, to the best of our knowledge, the analyze the interaction between the two aspects, let alone to analyze how developers can address them in the design process.

Therefore we decided, based also on previous endeavours and activities with older adults in our LivingLab, to use a case study of an intergenerational design team during a Hackathon event conducted at the Polish-Japanese Academy of Information Technology in Warsaw, Poland (PJAIT). In this study we addressed the issue of the connection between on-line security and real life safety in the design process by using participatory design approach. By describing the design process and group dynamics, as well as the final product, we wish to open further discussion, and provide some deeper insight into a topic that, although important in lieu of the aforementioned trend in stronger connection between the on-line and off-line, was not yet fully addressed by the research community. In this paper we provide more comprehensive approach to the security of  older adults based on the concept described in case study section which we previously briefly outlined in our conference report \cite{balcerzak2017press}.

The rest of the paper is organized as follows. In the \textbf{Related work} section, we provide a comprehensive description of findings in regards to on-line security and real life safety of older adults, as well as security design in software engineering, as well as participatory design, and older adults motivation for taking part in design processes and similar tasks. In the section  \textbf{LivingLab insight} we describe a broader context of their empirical studies and research activities with older adults in PJAIT LivingLab relevant to the above mentioned topics. The \textbf{Case study} section present in detail the design session conducted during the Hackathon, the specific features of the application created by the intergenerational team, as well as the design process itself. The \textbf{Conclusions and Future Work} section details how the observations made during the design session interact with the findings related to the topics of on-line security and real life safety. 

\section{Related works}\label{related_work}

\subsection{Older adults and on-line security}

According to studies done by \cite{grimes2010older} older adults are, on average less knowledgeable and aware of on-line risks than younger adults. Also, as shown by \cite{adams2005psychological}, older adults show limited trust to on-line technology, however this effect diminishes with usage of ICT. Another dimension crucial with on-line security is trust. Studies done by \cite{mccloskey2006importance} as well as \cite{grimes2010older} show that trust among older adults is not affected by age, and relies more on experience that comes with the use of ICT. Our previous study on location-based game with mobile ICT technology \cite{kopec2017location} which we present briefly in next section also showed the connection between direct ICT usage and self-confidence of older adults in context of mobile technologies. Therefore, the design of applications that address the issue of trust is important. The topic of safety, also plays a role in research related to applying smart house technologies as tools for enabling older adult independence\cite{peek2017can}.
A topic also related to on-line safety is the issue using data and text mining techniques for the detection f scamming and the credibility of on-line material\cite{Wawer:2014:PWC:2567948.2579000}\cite{Jankowski-Lorek:2014:PCW:2639968.2640074}.

\subsection{Older adults and real life safety}
For the sake of this paper, when reviewing the real-life hazards faced by older adults, the authors will focus on a specific aspects of safety, which in the literature is refereed to as neighborhood safety. This refers to risk related to social interaction in the nearest environment of the older adult. As the comprehensive literature review conducted by \cite{forjuoh2016172} suggests, five main themes constitute the overall framework of neighborhood safety. These are: general neighborhood safety; crime-related safety; traffic-related safety; fall-related safety; and proxies for safety (e.g., vandalism, graffiti).
This literature review provides also a deep and comprehensive review of what main themes are the object of focus among researchers interested in real life hazards and their perception by older adults. It is noteworthy, that in the studies used in the review, main focus was put on health issues and their relation with the feeling of safety. This is a theme which appears in works such as those done by \cite{won2016neighborhood}, where the authors show that physical health is a correlate with the older adults sense of safety. Another interesting field of study is related to older adults emotion in relation with surrounding environment. We conducted a preliminary study on mapping senior citizens' emotions with urban space \cite{nielek2017emotions}, which is also connected with our previous work, namely the mentioned  mobile game \cite{kopec2017location} which was related to the topics of well-being and happiness of older adults. This issue is a vital subject of many other interesting research, i.e a longitudinal study on a sample of over 10,000 older adults that  pointed out that perception of one's psychological well being affects the older adults feeling of safety \cite{choi2017perceived}.
To the best of the authors knowledge, little focus was placed on interaction between thus defined levels and perceptions of neighborhood safety and use of mobile devices and apps. Therefore, besides our previous endeavours mentioned above, we decided also to pursue this topic in exploratory in-depth interviews with older adults form our LivingLab, which are elaborated in the next section.

\subsection{User participation and interaction}
Fundamental idea relevant for this case is connected with a concept of user-center design, as a part of general idea of human focused approach related to another important idea: participatory design, sometimes called co-design. While there are different origins for the two latter terms  they actually refer to the same idea of bottom-up approach which is widely used besides software engineering from architecture and landscape design to healthcare industry.\cite{szebeko2010co} All those concepts put human in the center of designing process, however, there is a small, but significant distinction between user-center design and participatory design: the former refers the process of designing FOR users, while the latter WITH users.\cite{sanders2002user,sanders2008co} From the point of view of this study especially important are concepts that are related to user competences and empowerment provided e.g. by Ladner \cite{ladner2015design}.

Another case is the work done by \cite{xie2012connecting}, describing a cooperation between seniors and preschool children in a design task. An interesting observation made in this research was the fact that both groups needed equally time together, and time in separation in order to function properly. The broader context for these topics is covered by the contact theory, a widely recognized psychological concept coined by Allport \cite{allport1979nature} and developed for many years by other, e.g. Pettigrew.\cite{pettigrew1998intergroup} According to the theory  the problem of intergroup stereotypes can be faced by intergroup contact. However, there is several condition for optimal intergroup contact, but many studies proved that intergroup contact is worthwhile approach since it typically reduces intergroup prejudice.\cite{pettigrew2006meta} We had also explored the topic of intergroup interaction in our various studies including mentioned intergenerational location-based game and the hackathon based on previous works and tools i.e. by Rosencranz \cite{rosencranz1969factor}.

\subsection{Volunteering of older adults}

The effects of volunteering has on older adults is a developing field of study within various disciplines. The consensus that can be reached throughout various studies, is that participation in volunteer activity has many positive effects on the elderly. \cite{lum2005effects} claim that older adults who frequently volunteer in various activities, tend to have improved physical and mental health, compared to those who do not participate in volunteering. The work of \cite{morrow2003effects} extends this correlation to well-being (with volunteering being correlated with higher levels of well-being) this is also reinforced by \cite{greenfield2004formal}and \cite{hao2008productive}. This is crucial since studies also show that in some regards, elders are more likely to be engaged in volunteer activity \cite{morrow2010volunteering}.

Motivation for volunteering among older adults is also important. In research done by \cite{itoko2014involving} involving the comparison of older and younger adults when participating in a crowd sourcing task of proof reading texts in Japanese, it was shown that older adults where successfully motivated by the use of gamification techniques within the task, which means according to the most widespread definition of gamification by Deterding \textit{the use of game design elements in non-game contexts}.\cite{deterding2011game} However, a similar task, conducted by \cite{brewerwould} on a group of American seniors showed a slightly different pattern, where older adults were bored by the task, and did not comprehend its' significance. Nevertheless gamification is strictly connected with motivation and therefore inevitably leads to the psychological context, e.g. Zichermann claims that \textit{gamification is 75 percent of psychology and 25 percent of technology}. \cite{zichermann2011gamification} At this point it is worth mentioning that according to the latest reviews more and more solutions are based on solid psychological theories and frameworks. \cite{mora2015literature} One of the most important theoretical approach in the field of motivation is self-determination theory (SDT) developed by Ryan and Deci. \cite{ryan2000self} It is based on subtheories formerly developed by the authors of SDT: cognitive evaluation theory (CET) related to the intristic motivation and organismic integration theory (OIT) related to extrinsic motivation. The theory was proven to be effective to the elderly as well, e.g. by Vallerand .\cite{vallerand1989motivation} We had explored the topic of older adults volunteering and motivation in our research i.e. Wikipedia content co-creation. \cite{nielek2017turned}  

\section{LivingLab PJAIT insights} \label{livinglab}
In this section we provide some insights from our experience with older adults within Polish-Japanese Academy of Information Technology LivingLab. Further details of our LivingLab, it's origin and design as well as older adults activities are provided in separate description \cite{kopec2017livinglab}.

\subsection{About LivingLab PJAIT}
The term \textit{Living Lab} was coined by William Mitchell from MIT \cite{niitamo2006state} and was used to refer to the real environment, like a home or urban space, where routines and everyday life interactions of users and new technology can be observed and recorded to foster the process of designing new useful and acceptable products and services. The idea of LivingLab is therefore inherently coupled with broad concept human-centered approach described in previous section.\\
LivingLab at the Polish-Japanese Academy of Information Technology is a long-term framework project, whose goals are related to social inclusion and active engagement of the elderly in social life by facilitating the development of ICT literacy among them as well as creating an active community of stakeholders who are both the beneficiaries and enablers of research into their problems. 
This framework project has been established in a long-term partnership with the City of Warsaw.\\

Throughout recent years we have organized a number of activities for older adults focused on various research areas relevant to the topics of this article as we mentioned in related work section. In particular during an intergenerational location based game older adults performed various everyday mobile ICT tasks, such as connecting to Wi-Fi, browsing the information, scanning QR codes or taking panoramic photos with the assistance of their teammates. On the other hand, the younger participants benefited from the background knowledge about the city and its history of the elderly. Thus a positive bi-directional intergenerational interaction was observed alongside positive older adults self-awareness of the technology in context of physical activities and well-being. In other LivingLab activities, including on-line courses and crowdsourcing tasks as well as real life workshops and activities i.e. devoted to both application and content co-creation we also noticed that security issues are important but a bit vague area for older adults. Based on the literature review we decided to conduct a more in-depth qualitative research described in next subsection.

\subsection{Older adults and security}
Based on literature and observations from LivingLab activities, workshops and consultations we decided to obtain a more in-depth insight in the topic, including both relevant perspectives: on-line security and real life safety.

We used individual in-depth interviews in order to extract additional security insights. In total, we conducted four such interviews with older adults from our LivingLab aged 65+, two less technology advanced (female participants P1 and P2) and a pair of more advanced older adults (female P3 and male P4). To obtain more in-depth information from them we decided to conduct individual semi-structured interviews related to several topics: Internet and mobile application usage, endeavours towards on-line security and real life safety alongside with perception of potential interaction between those two realms.

Based on interviews we figured three different strategies toward security: caution, separation and awareness. While awareness is related to the more technologically advanced older adults, the two former strategies are interesting since they are represented by older adults with lower ICT literacy. Generally speaking the caution strategy represented by P1 was based on carefulness in both real life and on-line activities while separation strategy represented by P2 was based on strong conviction on non-transition between virtual and real realm. In particular P2 claimed that \textit{these two worlds should not be comined}.

First we asked participants about their Internet and mobile experience. There are three major areas of their on-line interests: 
\begin{itemize}
\item doing everyday tasks, like paying the bills
\item keeping in touch with family and friends
\item source of information.
\end{itemize}
The latter was the most extensive category and included various topics from health and medical issues (i.e. drugs and food ingredients, dietary information), through transportation to cultural and political news. In this context P2 claimed that \textit{The world has gone so far that it is difficult to live without the Internet these days}.

In order to obtain insight about on-line security we asked the participants about securing themselves in virtual space. In this context participants were aware of anti-virus protection as well trust issues, identity theft and identity verification (the need of verification the identity of on-line entities i.e. shops and companies). However, here we observed the major difference in separate-world approach. P2 explicitly stated, that there is no direct transition between the virtual and real realms. This was directly connected with careless websurfing. Moreover P2 wasn't afraid of her identity theft based on the claim \textit{I am no one special, I am an ordinary Smith}. 

We also asked about the safeness endeavor made by the participant in their everyday life. Besides personal physical safety measures like traffic safety, observation of the surroundings and other persons they pointed out two major areas connected with virtual space: finance and health. However they cannot establish the connection between the two realms. In particular, besides techniques of safe cash carrying (P1: \textit{one can deposit the cash in various piece of garment}) they stated that in general they use credit cards instead of cash and they deposit money in bank. In reference to health, besides physical activity they pointed out healthy diet and food and medical ingredients verification. 

As we can see some factors from virtual space activities can be directly mapped and connected with the real world. However, it was difficult to the participants to find the connection by themselves. More technologically advanced older adults (P3 and P4), were more aware of two realms i.e. bank account protection and identity theft alongside with interface as a vital concept between virtual and real world. On the other hand older adults with lower ICT literacy did not found themselves the connection between real life safety and increasing on-line security and vice versa. Surprisingly even though they provided a number of proper examples and behaviors from both realms they failed to find themselves spontaneously the connection. This leads to the conclusion that inevitably there is a room for designers to employ a participatory design approach in order to obtain insights from older adults not only to better understand their need but also to foster the process of idea development in order to find better connection between on-line and real life habits and safety. Because usually there is a generation gap between software development teams and end-users in case of older adults application we also decided to obtain a deeper insight into the dynamics of such collaboration, described in next section, based on our experience in the field of intergenerational interaction.

\section{Case Study}\label{design}
\subsection{Case study context}
The case study, involving the intergenerational developer team, was observed during a DEVmuster Hackathon organized in march 2016 in The Polish-Japanese Academy for Information Technology in Warsaw Poland, during which older adults and students of the academy had an opportunity to cooperate in designing application that would address the needs of older adults. The team consisted of 4 males, all in their twenties, who were students of the academy, and two older adults, a male, and a female, who were participants of the PJAIT LivingLab, a framework presented in previous section, which involves volunteers wishing to take part in various project aimed at activizing older adults with the use of ICT.

The team formed during the first hours of the event, during which ideas for the potential app were discussed. Upon reviewing the opinions of the older adults, the students decided to change their original idea of an application, in to an app designed for exchange of favor between volunteers and older adults. The name 'F1' was chosen, based on the function key for calling the help menu.

\subsection{Platform architecture}

The F1 platform was designed as client-server model and requires access to the Internet for the proper functioning. It might be a serious limitation for less developed and less populated countries but as the platform is intended to be deployed in Poland we decided to sacrifice versatility for simplicity. 

Access to the system is possible either through a web site or a dedicated mobile application. Although both ways provide the same functionality there are also substantial differences. Mobile application was designed to be most convenient for people offering support. Web site is more focused on posting requests for help. The reason for this differentiation is that mobile application will be more frequently used by younger volunteers and web browsers are better suited for older adults. In the case described in this paper senior participants were more familiar with traditional desktop or mobile computers than with smartphones, mainly due to their professional background and prior LivingLab activities i.e. computer workshop organized by the City of Warsaw. Moreover computer web browsers can be more suitable for people with certain disabilities -- e.g. visually impaired or with limited hand dexterity. On the other hand mobile devices are becoming more and more widespread and the adoption of touch interface by the elderly can be faster and more effective than traditional computer interfaces (e.g. observed in our previous research studies). Thus in final product the application mode (web site or mobile app) is intended to be freely interchangeable at any time by the user.

\subsection{Functionality}
\begin{figure}[p]
\centering
	\includegraphics[scale=0.35]{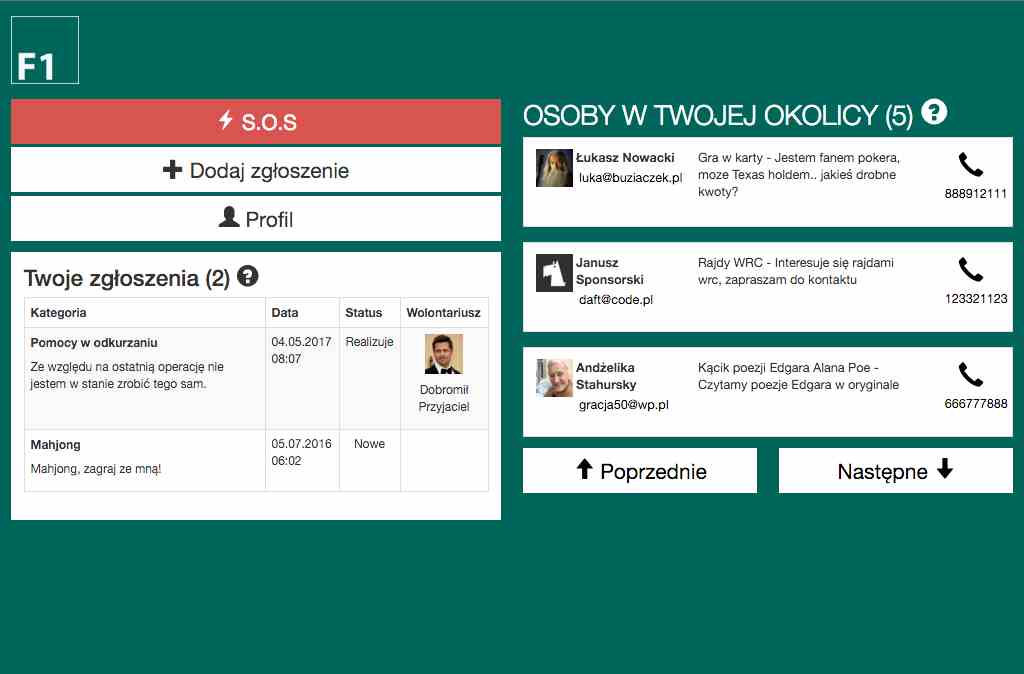}
	\caption{Main screen of web-based application focused mostly on people requiring assistance.}
    \label{F_main}
\end{figure}

As presented in figure \ref{F_main} 
The web based application contains all the key information and functions important for the user searching for the help of a volunteer. The screen informs the user about his or her previous favor requests, as well as the location of other users in the area. There is also an S.O.S button which can be used in case of emergency.

\begin{figure}[!tbp]
  \centering
  \subfloat[Welcome and login screen.]{\includegraphics[width=0.3\textwidth]{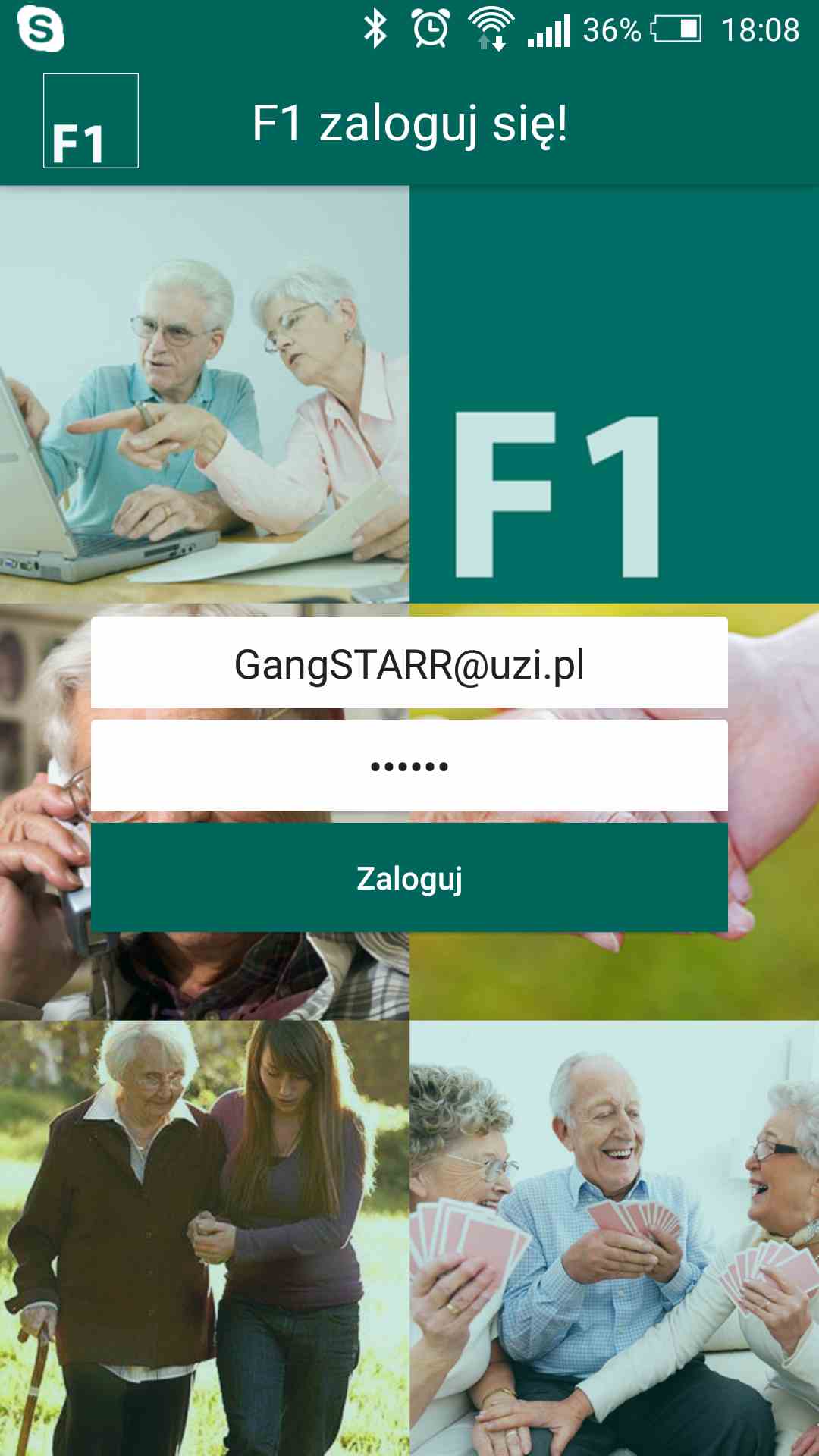}\label{fig:f3}}
  \hfill
  \subfloat[Screen presents requests for assists in neighborhood on the map.]{\includegraphics[width=0.3\textwidth]{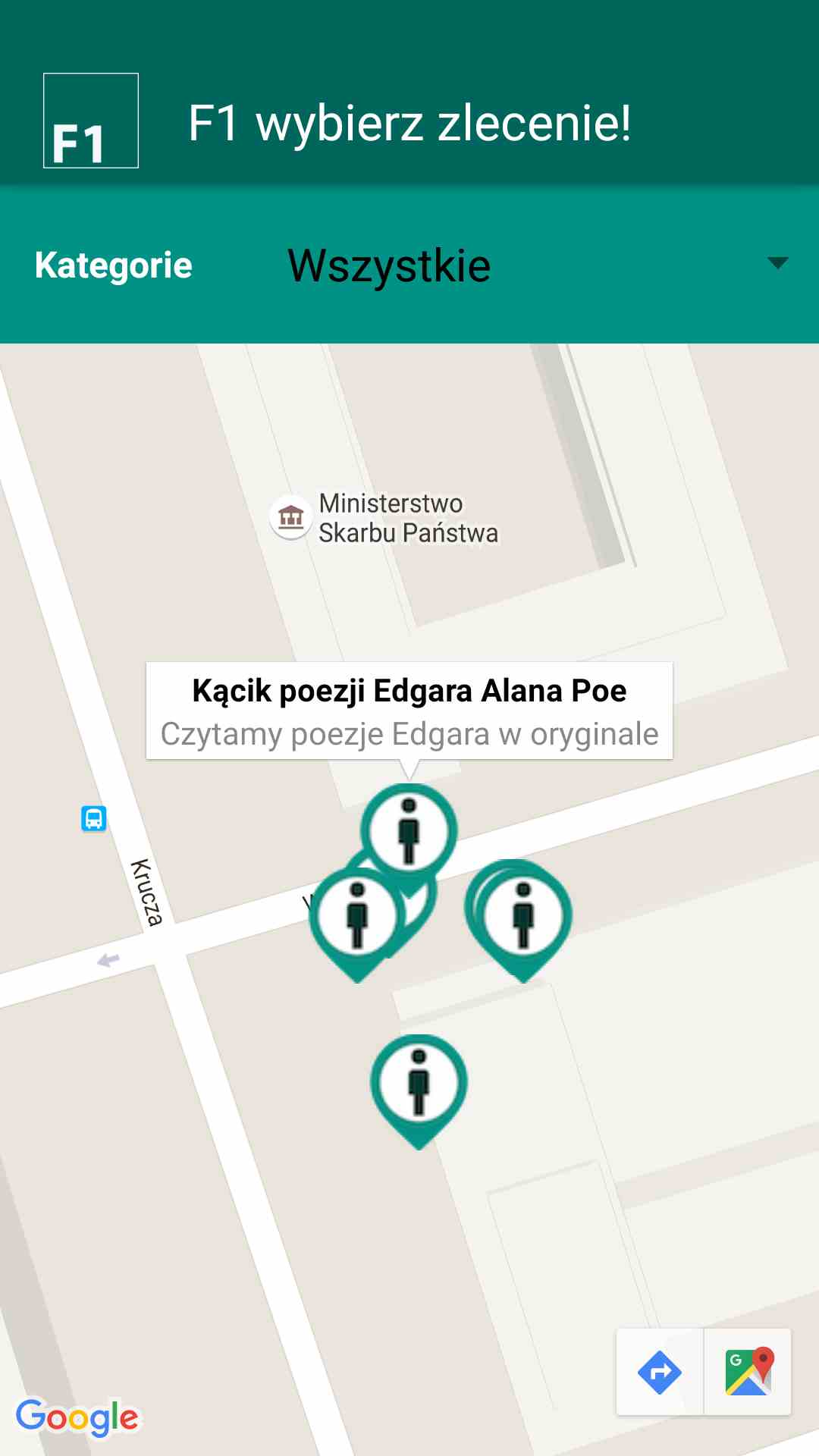}\label{fig:f2}}
  \hfill
  \subfloat[Detailed information about request for assist.]{\includegraphics[width=0.3\textwidth]{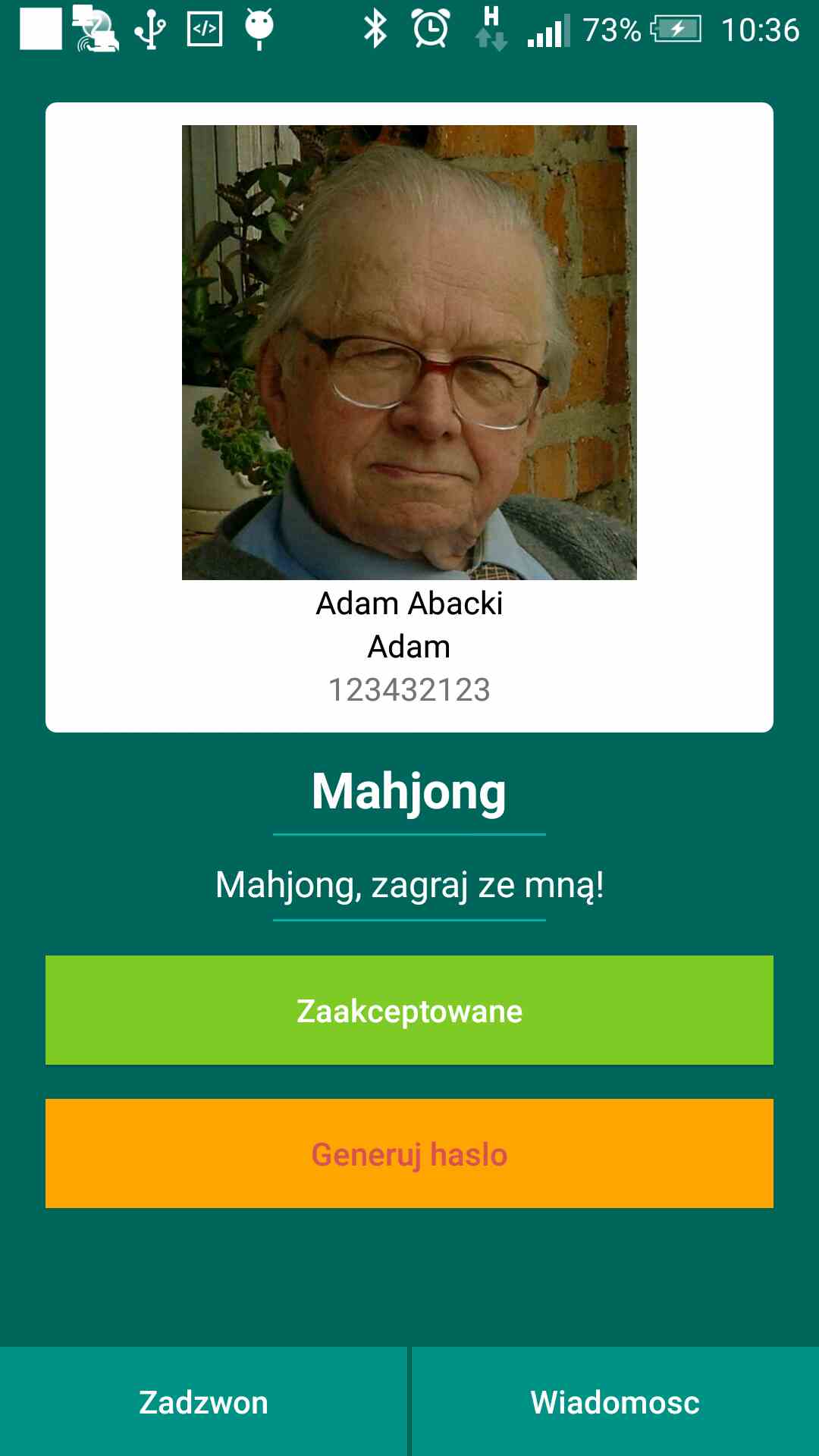}\label{fig:f1}}
  \caption{Mobile application dedicated for those who offer assistance.}
\end{figure}

The mobile application view, used by the potential volunteer is shown in figure\ref{fig:f1}. When using the mobile view, the user can view a map of the nearest area where favor requests are marked. When the user selects a favor a brief description of the favor and the requesting user is provided.

\begin{figure}[!ht]
\centering
	\includegraphics[scale=0.15]{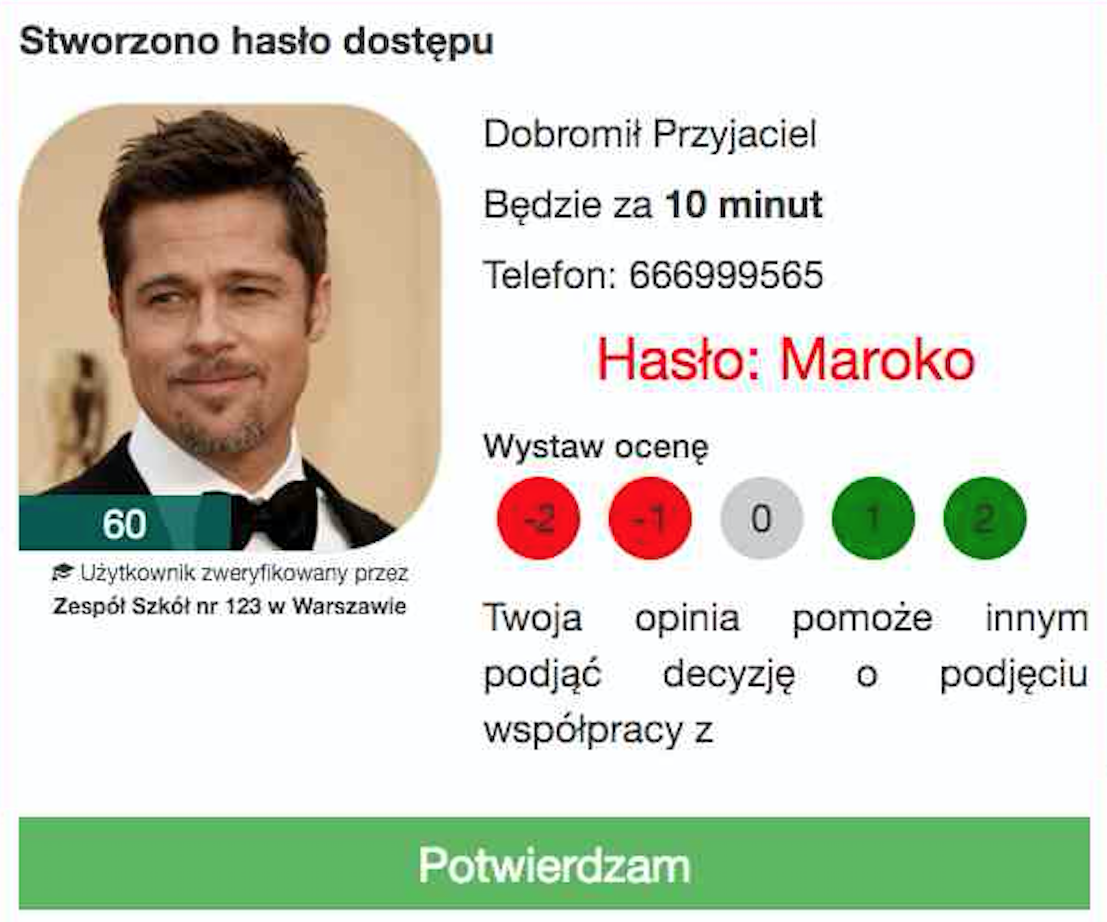}
	\caption{Pop-up window displaying the confirmation details for selected volunteer}
    \label{F_popup}
\end{figure}

\subsection{Security related features}

Another direct benefit from applying participatory design approach refers to security issues. During the pre-design phase, older adults voiced their concerns related to user security, taking into account threats for both parties of the  process. The main areas in which they stated older adults require aid when it comes to security were related with minimizing the risks of coming in contact with fake profiles, or malicious users, as well as dealing with problems of potential identity theft. The older adults involved in the project also stressed that the platform should be able to aid the older users when dealing with emergencies when swift help is needed. These issues were addressed by applying the following solutions into the design of the platform.

\subsubsection{Trusted profiles}

Sign up is free for all and requires only a valid email account. Lowering the barrier makes system more user-friendly but also prone to malicious users. Discussions with prospect users during design phase reveal that the elderly are afraid of letting unknown people visit their apartments. To address this problem a voluntary procedure for confirming profiles was implemented. User profile might be confirmed by external  organizations that are trustworthy: e.g. schools or local NGOs. Confirmation of the verified status is visible for everyone next to profile picture -- see fig. \ref{F_popup}.

\subsubsection{Challenge-response authentication}

Next to the threat of fake or malicious profiles mentioned in previous section yet another problem was identified by participatory approach. Even if identity of volunteer is confirmed on the platform still we need a way to confirm it in real world when volunteer is knocking to the door of senior's apartment. This is the situation when digital system should face analogue world and bottom-up approach proved to be helpful once again. Therefore, a standard challenge-response authentication has been adapted and implemented. The platform generates two keywords for both users. Elderly should ask about the right password before letting someone in. Passwords are randomly selected from a subset of polish words to make them easy to remember and dictation by entry phone.

\subsubsection{Reputation score}

Limiting the amount of frauds is crucial for assuring wide acceptance of the platform but it is not enough. Next to deliberate and planed frauds we can also see a bad quality service. Therefore, the platform contains a reputation management system. Every agreed and conducted service can be evaluated on Likert-type scale. To make scale easier to understand for users first two grades are red, neutral score is gray and the two positive levels are green. Sum of all evaluation for give user are displayed next to the picture -- see fig. \ref{F_popup}. 

\subsubsection{Emergency button}

In real life exhaustive list of risks and threats is impossible to complete. Therefore, an emergency button has been added to the F1 platform. 

\section{Conclusions and future work}\label{conclusion}

As it was presented in the \textbf{Related Work} section and in our study, the issues of on-line security and real life safety faced by older adults are quite diverse. The span a wide set of threats ranging from health issues, to crime related issues, and problems connected with the spread of malware and spam, and even though older adults can point out a variety of those issues, usually it is difficult for them to spontaneously find the connection between the two realms. However, in our case study we showed that participatory design approach can benefit both younger tech-minded developers and older adults as end-users.

The observations made during the presented case study show, that the key for creating an on-line solution, that would not only aid them in their daily lives, but also but be resilient to safety and security issues mentioned in the literature, is not necessarily to address all potential dangerous scenarios. From the description of the app created through a participatory design within an intergenerational team, where older adults, were not only final users, but also active team members, one can see that the focal point for addressing issues of safety, both on-line and in real life, is modeling of trust within the user base of the application. Use of such elements like a reputation system, two step password verification between users who decided to exchange favors, as well as external confirmation of users provides a wide array of possibilities for encouraging the development of trust between older adults and younger volunteers. This corroborates the general ramifications of the intergroup contact theory, however it extends its scope beyond intergroup stereotyping into the field of limiting the feeling of insecurity among the group of older adults.

It is also important to notice that the participatory approach utilized by the team, allowed to overcome the aforementioned issue of understanding the link between real life safety and on-line security. In light of the results from the the interviews conducted within the framework of the LivingLab, where older adults had difficulties with connecting the two aspects of safety, the outcome of the team design process yields great promise.

The case study presented here, of course, serves mostly as a jumping-off point for further considerations in a topic that, although important, is not, in the authors opinion, amply researched. The findings made in this paper show, that by use of a participatory design framework, it is possible not only to address general issues of on-line security, but also, to create new methods of thinking about user safety outside of the paradigm of software engineering. With the increasing role of mobile technologies in real life situation, these initial exploratory findings offer interesting option for improving the initial design process of mobile applications.

In their future work, we plan to further explore this interesting emerging field of study. With the observation made in this paper being mostly exploratory in nature, it seems fitting to conduct a set of more methodologically rigid tests with the aim of verifying, what form of participatory design can further improve the process of addressing the threat on-line and real life threats faced by end users of an application.

\section{Acknowledgments.}

This project has received funding from the European Union’s Horizon 2020 research and innovation programme under the Marie Skłodowska-Curie grant agreement No 690962.

%
%

\bibliographystyle{abbrv}
\bibliography{literature}







\end{document}